\documentclass[apjl]{emulateapj-rtx4}

\usepackage{hyperref}
\usepackage{natbib}
\shorttitle{{\textit{Spitzer}} imaging of {\textit{Herschel}}-ATLAS lenses}
\shortauthors{Hopwood et al.}

\begin{document}



\title{{\textit{Spitzer}} Imaging of {\textit{Herschel}}-ATLAS Gravitationally Lensed
  Submillimeter Sources}

\author{R. Hopwood\altaffilmark{1}, 
J. Wardlow\altaffilmark{2},
A. Cooray\altaffilmark{2}, 
A. A. Khostovan\altaffilmark{2}, 
S. Kim\altaffilmark{2}, 
M. Negrello\altaffilmark{1}, 
E. da Cunha\altaffilmark{3},  
D. Burgarella\altaffilmark{4},
I. Aretxaga\altaffilmark{5},
R. Auld\altaffilmark{6},
M. Baes\altaffilmark{7},
E. Barton\altaffilmark{2},
F. Bertoldi\altaffilmark{8},
D. G. Bonfield\altaffilmark{9}, 
R. Blundell\altaffilmark{10}, 
S. Buttiglione\altaffilmark{11},
A. Cava\altaffilmark{12}$^{,}$\altaffilmark{13},
D. L. Clements\altaffilmark{14},
J. Cooke\altaffilmark{3}$^{,}$\altaffilmark{15},
H. Dannerbauer\altaffilmark{16},
A. Dariush\altaffilmark{6}$^{,}$\altaffilmark{35}, 
G. de Zotti\altaffilmark{11}$^{,}$\altaffilmark{17},
J. Dunlop\altaffilmark{18},
L. Dunne\altaffilmark{19},
S. Dye\altaffilmark{6},
S. Eales\altaffilmark{6},
J. Fritz\altaffilmark{7},
D. Frayer\altaffilmark{20},
M. A. Gurwell\altaffilmark{10},
D. H. Hughes\altaffilmark{5},
E. Ibar\altaffilmark{21},
R. J. Ivison\altaffilmark{18}$^{,}$\altaffilmark{21},
M. J. Jarvis\altaffilmark{9},
G. Lagache\altaffilmark{23}$^{,}$\altaffilmark{24},
L. Leeuw\altaffilmark{25}$^{,}$\altaffilmark{26}, 
S. Maddox\altaffilmark{19},
M. J. Micha{\l}owski\altaffilmark{18},
A. Omont\altaffilmark{27},
E. Pascale\altaffilmark{6},
M. Pohlen\altaffilmark{6},
E. Rigby\altaffilmark{19},
G. Rodighiero\altaffilmark{28},
D. Scott\altaffilmark{29},
S. Serjeant\altaffilmark{1},
I. Smail\altaffilmark{30},
D. J. B. Smith\altaffilmark{19},
P. Temi\altaffilmark{31},
M. A.Thompson\altaffilmark{9},
I. Valtchanov\altaffilmark{32},
P. van der Werf\altaffilmark{18}$^{,}$\altaffilmark{23},
A. Verma\altaffilmark{34},
J. D. Vieira\altaffilmark{15}
}
\altaffiltext{1}{Department of Physics and Astronomy, The Open University, 
Milton Keynes, MK7 6AA, UK}
\altaffiltext{2}{Department of Physics and Astronomy, University of California, Irvine, 
CA 92697, USA}
\altaffiltext{3}{Department of Physics, University of Crete, Heraklion, Greece}
\altaffiltext{4}{Laboratoire d'Astrophysique de Marseille, OAMP, CNRS, Aix-Marseille Universit\'e, France}
\altaffiltext{5}{Instituto Nacional de Astrof\'{\i}sica, \'Optica y Electr\'onica,  
Aptdo. Postal 51 y 216, 72000 Puebla, Mexico}
\altaffiltext{6}{School of Physics and Astronomy, Cardiff University,
Cardiff, CF24 3AA, UK}
\altaffiltext{7}{Sterrenkundig Observatorium, Universiteit Gent,
 Krijgslaan 281 S9, B-9000 Gent, Belgium}
\altaffiltext{8}{Argenlander Institute for Astronomy, University of Bonn, 
 Auf dem Hugel 71, 53121 Bonn, Germany}
\altaffiltext{9}{Centre for Astrophysics Research, Science and Technology Research Centre, 
 University of Hertfordshire, Herts AL10 9AB, UK}
\altaffiltext{10}{Harvard-Smithsonian Center for Astrophysics, 
Cambridge, MA 02138, USA}
\altaffiltext{11}{INAF, Osservatorio Astronomico di Padova,
 Vicolo Osservatorio 5, I-35122 Padova, Italy}
\altaffiltext{12}{Instituto de Astrofisica de Canarias,
    C/VÌa Lactea s/n, E-38200 La Laguna, Spain}
\altaffiltext{13}{Departamento de Astrofisica, Universidad de La Laguna (ULL), 
    E-38205 La Laguna, Tenerife, Spain}
\altaffiltext{14}{Astrophysics Group, Physics Department, Imperial College 
 London, Prince Consort Road, London SW7 2AZ, UK}
\altaffiltext{15}{California Institute of Technology, 
    MS 249-17, 1216 E. California Blvd., Pasadena, CA 91125, USA }
\altaffiltext{16}{Laboratoire AIM  Paris Saclay, CEA-CNRS-Universit\'{e}, Irfu/Service d'Astrophysique, CEA Saclay, Orme de Merisiers, 91191 Gif-sur-Yvette Cedex, France}
%
\altaffiltext{17}{Scuola Internazionale Superiore di Studi Avanzati,
    Via Bonomea 265, I-34136 Trieste, Italy}
\altaffiltext{18}{SUPA, Institute for Astronomy, University of
  Edinburgh, Royal Observatory, Edinburgh, EH9 3HJ, UK}
%
\altaffiltext{19}{School of Physics and Astronomy, University of Nottingham,
 University Park, Nottingham NG7 2RD, UK}
%
%
%
%
\altaffiltext{20}{National Radio Astronomy Observatory
 P.O. Box 2, Green Bank, WV 24944, USA}
%
%
\altaffiltext{21}{UK Astronomy Technology Center, Royal Observatory Edinburgh,
 Edinburgh, EH9 3HJ, UK}
%
\altaffiltext{22}{Scottish Universities Physics Alliance, 
    University of Edinburgh, Royal Observatory, Edinburgh, EH9 3HJ, UK}
\altaffiltext{23}{Institut dÕAstrophysique Spatiale (IAS), Batiment 121, Universit\'{e} Paris Sud 11 91405 Orsay, France}
%
%
\altaffiltext{24}{CNRS (UMR8617), 91405 Orsay, France}
%
\altaffiltext{25}{Physics Department, University of Johannesburg,
    P.O. Box 524, Auckland Park 2006, South Africa }
\altaffiltext{26}{SETI Institute, Mountain View, CA, 94043, USA}
%
%
\altaffiltext{27}{Institut d'Astrophysique de Paris, Universit\'{e} Pierre et Marie Curie and CNRS,
 98 bis boulevard Arago, 75014 Paris, France}
%
%
%
%
\altaffiltext{28}{Dipartimento di Astronomia, Universita di Padova,
 Vicolo Osservatorio 2, I-35122 Padova, Italy}
%
\altaffiltext{29}{Department of Physics and Astronomy, University of British
  Columbia, Vancouver, BC V6T 1Z1, Canada}
%
%
%
\altaffiltext{30}{Institute for Computational Cosmology, Durham University, South Road,
Durham, DH1 3LE, UK}
\altaffiltext{31}{Astrophysics Branch, NASA Ames Research Center,
 Mail Stop 245-6, Moffett Field, CA 94035, USA }
%
%
\altaffiltext{32}{Herschel Science Centre, European 
 Space Agency, P.O. Box 78, 28691 Villanueva da la Ca\~nada, Madrid, Spain}
%
\altaffiltext{33}{Leiden Observatory, Leiden University, P.O. Box 9513, NL 2300 Leiden, The Netherlands}
\altaffiltext{34}{Oxford Astrophysics, Denys Wilkinson Building, University of Oxford,
 Keble Road, Oxford, OX1 3RH, UK}

%
%
  
\altaffiltext{35}{School of Astronomy, Institute for Research in Fundamental Sciences (IPM), PO Box 19395-5746, Tehran, Iran}

\begin{abstract}
  We present physical properties of two submillimeter selected
  gravitationally lensed sources, identified in the {\it Herschel}
  Astrophysical Terahertz Large Area Survey.  These submillimeter
  galaxies (SMGs) have flux densities $>$ 100$\,$mJy at 500$\,\mu$m,
  but are not visible in existing optical imaging. We fit light profiles to each
  component of the lensing systems in {\it Spitzer} IRAC 3.6 and
  4.5$\,\mu$m data and successfully disentangle the foreground lens
  from the background source in each case, providing important
  constraints on the spectral energy distributions (SEDs) of the
  background SMG at rest-frame optical-near-infrared wavelengths. The
  SED fits show that these two SMGs have high dust obscuration with
  $A_{\rm V} \sim 4-$ 5 and star formation rates of $\sim100$
  $M_{\odot}$yr$^{-1}$. They have low gas fractions and low
  dynamical masses compared to 850$\,\mu$m selected galaxies.
\end{abstract}


\section{Introduction}

Gravitational lensing is an invaluable astrophysical tool.  It can be
exploited to study galaxies beyond instrumental blank field sensitivities
and to constrain the total mass of galaxy systems, without regard for
its dark or luminous nature (see review by Treu 2010). 
The magnification of distant galaxies through gravitational lensing
enables the detailed study of sources that would otherwise be
undetectable. In the submillimeter regime this includes members of the
population of intrinsically faint galaxies that comprise a significant
fraction of the cosmic far-infrared background.

The {\it Herschel}\footnotemark[35] (Pilbratt et al.~2010) Astrophysical Large Area
Survey (H-ATLAS; Eales et al.~2010) is the largest open-time key project
currently being undertaken by {\it Herschel} and aims to survey 550
deg$^2$ and detect $> 300{,}000$ galaxies. A new methodology for selecting
gravitational lenses using wide-area submillimeter (sub-mm) surveys
(Blain 1996, Perrotta et al.~2002; Negrello et al.~2007) has been
tested for the first time during {\it Herschel}'s science demonstration
phase (SDP; Negrello et al.~2010, N10 hereafter). Candidates are
selected using a flux cut of 100$\,$mJy at 500$\,\mu$m, a limit based on
the steep number counts slope in the sub-mm (Negrello et
al.~2007). Above this flux limit only gravitationally lensed objects
and easily identifiable `contaminants' remain e.g. blazars and local
spiral galaxies. Initial results show that this selection method has
$\sim$100\% efficiency and should deliver a sample of hundreds of new
gravitational lenses in planned wide area {\it Herschel} surveys,
probing galaxies with intrinsic fluxes below the {\it Herschel}
confusion limit.

\footnotetext[35]{{\it Herschel}
    is an ESA space observatory with science instruments provided by
    European-led Principal Investigator consortia and with important
    participation from NASA.}

Five lens candidates in the SDP H-ATLAS data (Pascale et al.~2010;
Rigby et al.~2010; N10) were confirmed with spectroscopic redshifts
obtained via the detection of carbon monoxide (CO) emission lines (Lupu
et al.~2010; Frayer et al.~2011) of the background galaxies and optical 
spectra (Negrello et al. 2010) of the lens galaxies.  
In this Letter we study two of these gravitational lenses,
H-ATLAS J090311.6+003906 (SDP.81) and H-ATLAS
J091305.0-005343 (SDP.130), which are submillimeter galaxy (SMG) at redshifts of $z = 3.04$ and $z = 2.63$ being lensed by ellipticals at 
$z = 0.30$ and $z = 0.22$, respectively. Submillimeter Array (SMA) imaging 
reveals the sub-mm morphology, consistent with a lensing event, with multiple peaks
distributed around the position of the foreground elliptical galaxy
(N10). 
While no lensed background images were detectable in
optical imaging, the spectral energy distribution (SED) models suggest that
flux from the background sources should be detectable above
$\sim$10$\,\mu$Jy at near-IR wavelengths, from 2 to 5~$\,\mu$m (N10). At
these wavelengths, the emission from the foreground lenses, at
$z\sim0.25$, and the background SMGs becomes comparable.

In this Letter we present {\it Spitzer}/IRAC imaging, light
profile models, photometry and physical characteristics for SDP.81 and SDP.130
derived from SED fitting.  In the next section, we present a summary of
the {\it Spitzer} data, while in Section~3 we discuss modeling of the
light profile of the foreground lenses.  We perform SED modeling of
background SMGs and present results related to the properties of these
two sources in Section~4. Throughout the Letter we assume flat
$\Lambda$CDM cosmology with $\Omega_{\rm M}\,$=$\,0.3$ and $H_0\,=\,70$
km$\,$s$^{-1}\,$Mpc$^{-1}$.

\section{{\it Spitzer} IR and Keck Optical Imaging data}

The IR imaging data are part of {\it Spitzer} program 548 (PI:
A.\,Cooray), released for analysis on 2010 July 30.  For this
program, Infrared Array Camera (IRAC; Fazio et al. 2004) images of
SDP.81 and SDP.130 were taken at 3.6$\,\mu$m (Channel 1) and 4.5$\,\mu$m (Channel 2).
 Both lensing systems were imaged with a
36-position dither pattern and an exposure time of 30 s for each
frame to achieve an effective total exposure time of 1080 s.  The
rms depth reached is 0.3 and 0.4 $\mu$Jy in Channels 1 and 2,
respectively.

\begin{figure}[!htbp]
\begin{center}
\includegraphics[width=0.37\textwidth]{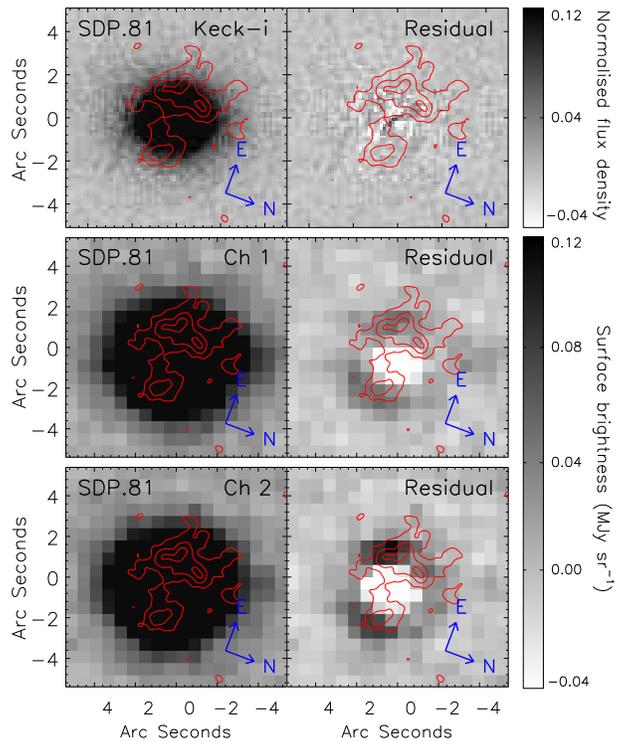}\hspace*{0.2cm}
\end{center}
\vspace{-0.4cm}
\caption{Keck {\it i}-band (top row), IRAC Channel 1 (middle row), and Channel
  2 (bottom row) postage stamp images for SDP.81. The
  corresponding residual (right column) is presented after subtraction of single de Vaucouleurs
  profiles modeled on the Keck {\it i}-band data. The Keck residuals show no
  signs of structure associable with the overlaid 880
  $\mu$m SMA signal-to-noise ratio (S/N) contours. For the IRAC residual there is
  structure that can be associated with the SMA contours, which
  suggests this is lensed structure of the respective background
  galaxy. The SMA contours are overlaid in steps of 3$\sigma$,
  8$\sigma$, 13$\sigma$ etc. SMA peaks with S/N $>$ 8 (four for ID81
  and two for ID130) have a maximum rms on the position of a point
  source of 0.1$''$. All postage stamps are plotted with the same pixel 
  intensity scaling after normalizing the Keck {\it i}-band data to the 
  dynamic pixel value range 
of the IRAC data. }
\label{single_profiles1}
\end{figure}

Corrected basic calibrated data pre-processed by the {\it Spitzer}
Science Center (SSC), using the standard pipeline version S18.18.0,
were spatially aligned, resampled, and combined into a mosaic image
using version 18.3.1 of the SSC's MOPEX software suite (Makovoz $\&$
Marleau 2005).  The IRAC mosaics have a resampled pixel size of 0.6$''$
and angular resolution of 2 to 2.5$''$.
For the work presented here, we also use the
IRAC point spread function (PSF; version 2010 April) file as provided
by the SSC
\footnote[36]{\url{http://ssc.spitzer.caltech.edu/irac/calibrationfiles/psfprf/}}.

\begin{figure}[!htbp]
\begin{center}
\includegraphics[width=0.37\textwidth]{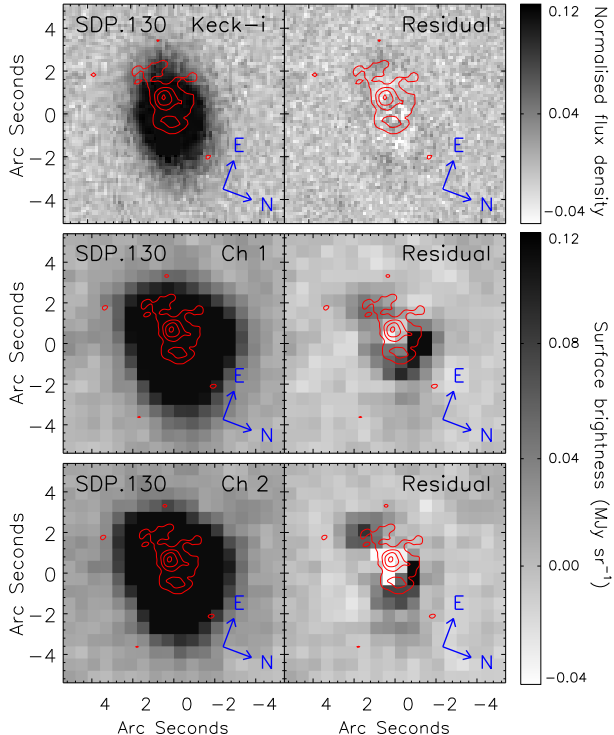}
\end{center}
\vspace{-0.4cm}
\caption{Keck {\it i}-band (top row), IRAC Channel 1 (middle row) and Channel
  2 (bottom row) postage stamp images for SDP.130. The
  corresponding residuals are presented after subtraction of single de Vaucouleurs
  profiles modeled on the Keck {\it i}-band data. Although the Keck residuals show no
  signs of significant structure, the subtracted IRAC data have
  residual structure that can be associated with the SMA contours, which
  suggests lensed structure of the respective background
  galaxy is present. The SMA contours are presented as in
  Figure~\ref{single_profiles1}. All postage stamps are plotted with the same pixel 
  intensity scaling after normalizing the Keck {\it i}-band data to the 
  dynamic pixel value range 
of the IRAC data. }
\label{single_profiles2}
\end{figure}

We also make use of optical images of SDP.81 and SDP.130 that were
acquired on 2010 March 10 using the dual-arm Low Resolution Imaging
Spectrometer (LRIS; Oke et al.~1995, McCarthy et al.~1998) on the Keck
I telescope and reported in N10 (see Figures~\ref{single_profiles1} and \ref{single_profiles2}).  Each
target received simultaneous 3$\,\times\,$110 s integrations with
the {\it g}-filter and 3$\,\times\,$60  integrations with the {\it i}-filter
using the blue and red arms of LRIS, respectively. A 20$''$ dither
pattern was employed to generate on-sky flat-field frames.  We performed
photometric calibration using 1 s {\it g}- and {\it i}-band observations
of bright stars in each field.  The data were reduced using IDL
routines and combined and analyzed using standard IRAF tasks; the
seeing FWHM of the final science exposures is $\sim$0.8$''$.

\begin{figure}[!htbp]
\begin{center}
\includegraphics[width=0.37\textwidth]{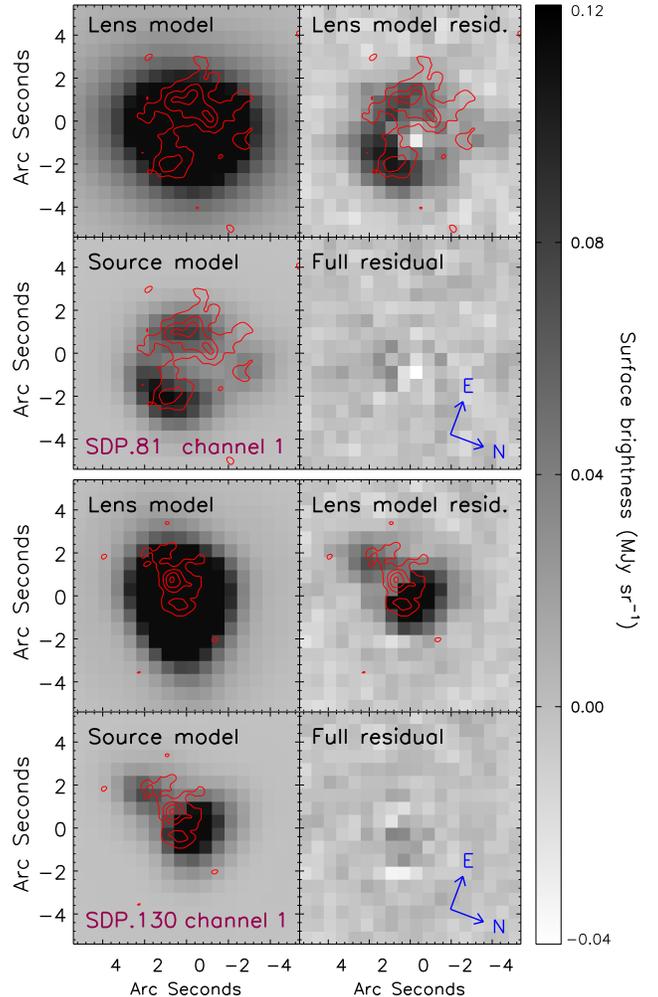}
\end{center}
\vspace{-1cm}
\caption{Multi-component light profile models for SDP.81 (top) and
  SDP.130 (bottom), for the IRAC Channel 1 data. Shown are the un-subtracted data, the
  lens model (a single S\'{e}rsic profile for SDP.81 and a S\'{e}rsic
  plus exponential disk profile for SDP.130 ), the
  residual after subtraction of the lens model profile only and the total model
  residual (full residual). The SMA contours are presented as in
  Figure~\ref{single_profiles1}. All data are plotted with the same pixel intensity scaling.}
\label{id81multi}
\end{figure}

\section{Modeling the lenses}

To construct models of the light profiles for each lensing system we
use GALFIT (Peng et al.~2002), which allows multiple profiles per
object and performs a simultaneous nonlinear minimization.  Prior to
fitting profiles to SDP.81 and SDP.130 in the IRAC data, de Vaucouleurs
profile (de Vaucouleurs~1948) models are constructed for the Keck
{\it i}-band imaging, to take advantage of the comparatively higher
resolution of these images, and the presence of only one component per
system i.e. the lens galaxy.

To look for any potentially lensed structure in the IRAC data, the 
Keck models are used to fit the IRAC Channels 1 and 2 data,
keeping the effective radius and ellipticity fixed, and using the
appropriate IRAC PSF $^{36}$.  On subtraction of the results, and in
comparison to the model subtracted Keck data and SMA contours, the
IRAC band residuals strongly suggest that a more complex structure, associated with
the background SMG, is present for both SDP.81 and SDP.130 (see
Figures~\ref{single_profiles1} and \ref{single_profiles2}).  

We verify that these residuals are significant, and not an artifact of
imperfections in the IRAC PSF, by comparing them with residuals derived
for three (non-lensing) elliptical-like galaxies in the same field, after
fitting them with single S\'{e}rsic profiles. For both the lensing
systems and the comparison ellipticals aperture flux ratios were determined
for the residual image and the corresponding un-subtracted data.
To consider only positive structure, pixel values $>\,2\,\sigma$ below
the local background were replaced with the median local sky value.
The SMG residuals were found to have flux ratios $\sim\,3-5$
greater than those for the random elliptical galaxies.

In order to more precisely disentangle the lens and background 
components we represent the flux from the lensed SMGs with 
S\'{e}rsic profiles. Peaks were identified in the IRAC single profile 
fit residual images, by fitting Gaussian profiles. We add three 
profiles for SDP.81 and two for SDP130, which all correspond 
to significant sub-mm contour peaks in the SMA data, within 0.8$''$.
N10 modeled the Keck {\it i} band with a S\'{e}rsic profile
plus an exponential disk, as this combination gave a marginally
improved $\chi^2$ over single profile models, and found that the
exponential component significantly contributes to the total profile
for SDP.130.  Therefore, a S\'{e}rsic plus exponential disk profile
is adopted for the SDP.130 lens and a single S\'{e}rsic profile for the
SDP.81 lens. 
To produce the final models for the lens and background galaxies, 
we refit the profile representing the lens galaxy and 
the additional S\'{e}rsic profiles simultaneously for each system with GALFIT.
These final model fits for the background component of SDP.81 show a
partial ``Einstein ring''-like morphological structure for the lensed
source (Figure.~\ref{id81multi}).  
For SDP.130, the lensed component is more compact and in close
proximity to the lens profile.  The resulting best fit profiles for
the background galaxies agree well with the SMA contours, and the
combined models subtract cleanly, suggesting successful lens/source
decoupling.

\begin{figure*}[!htbp]
\begin{center}
\plottwo{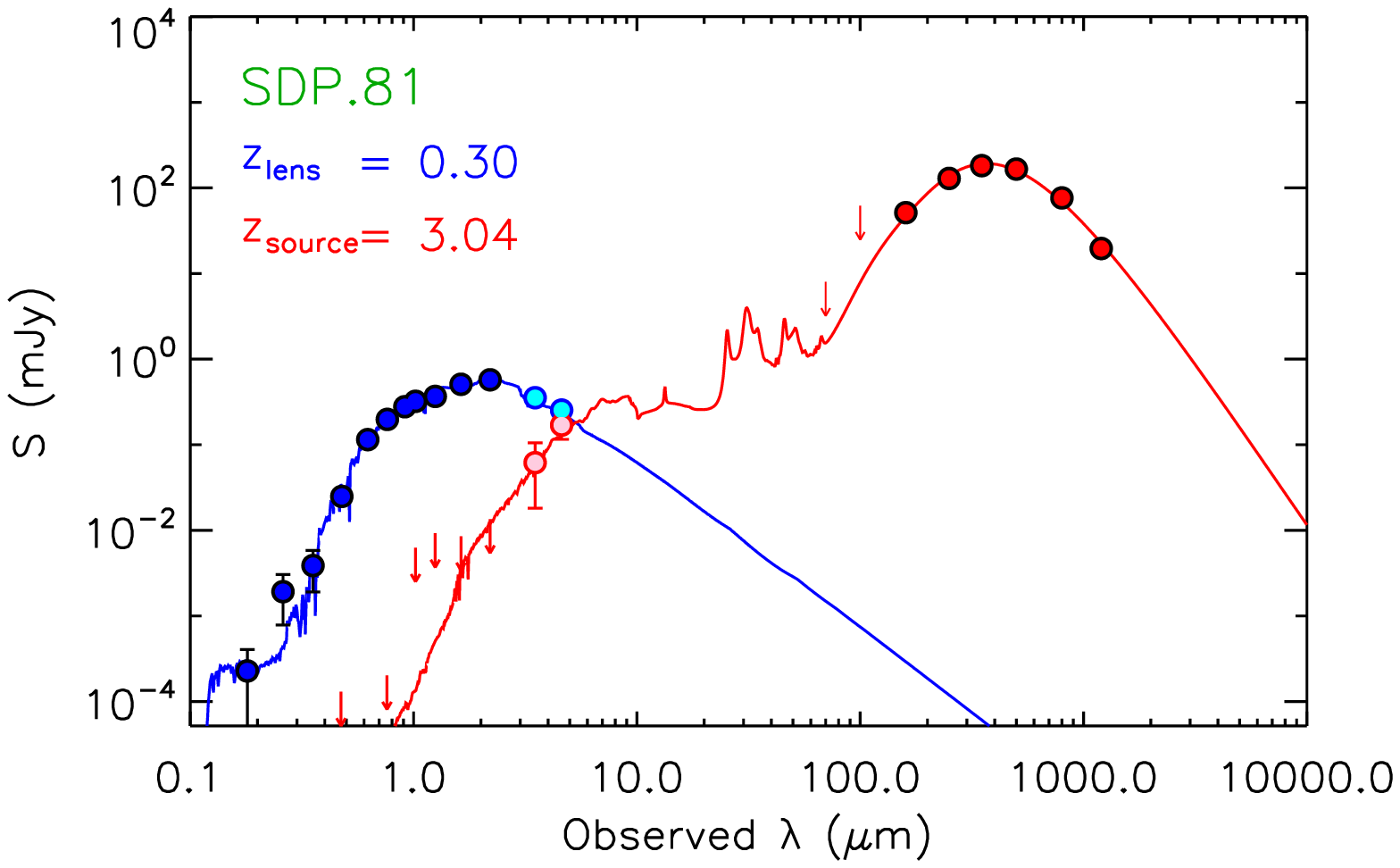}{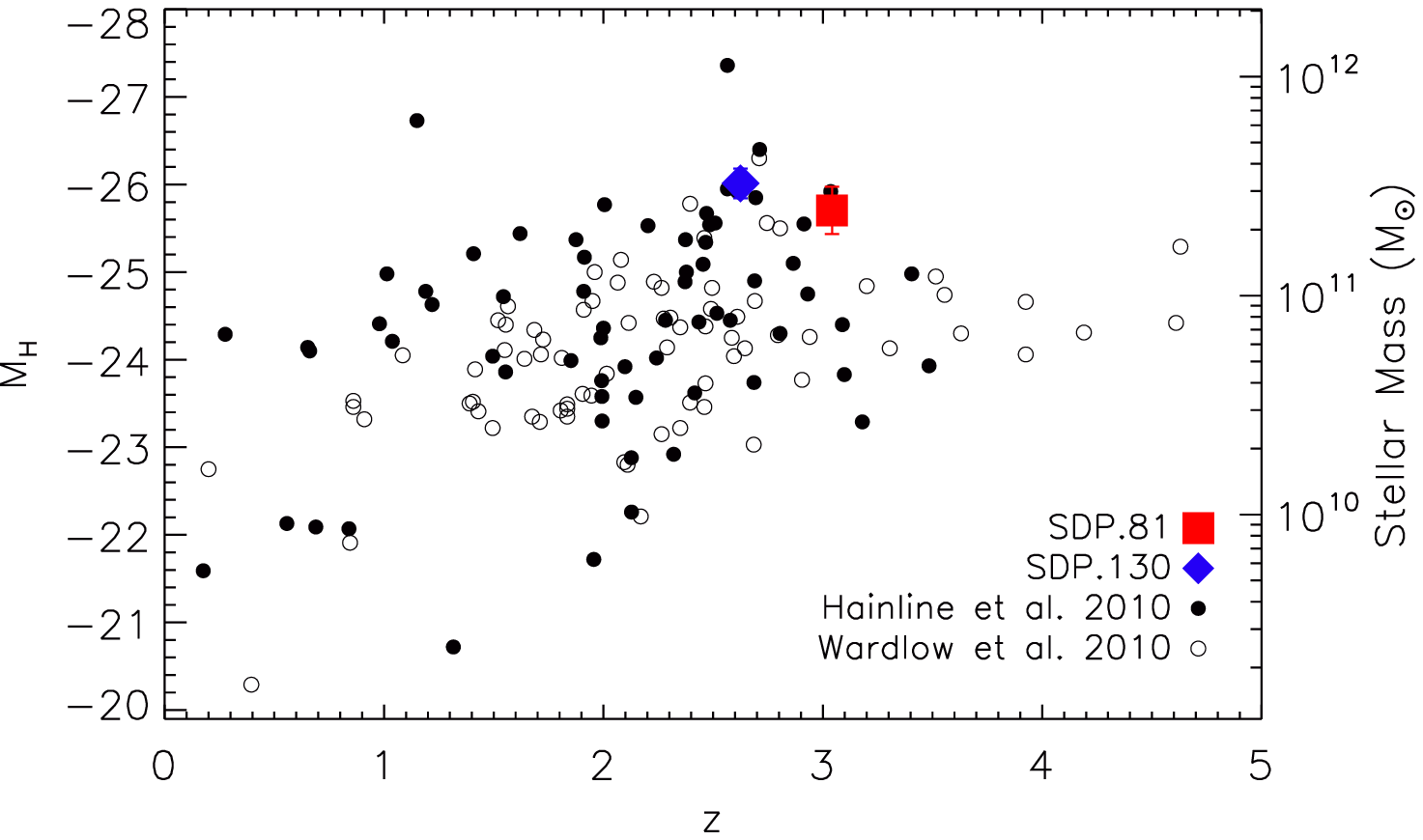}
\plottwo{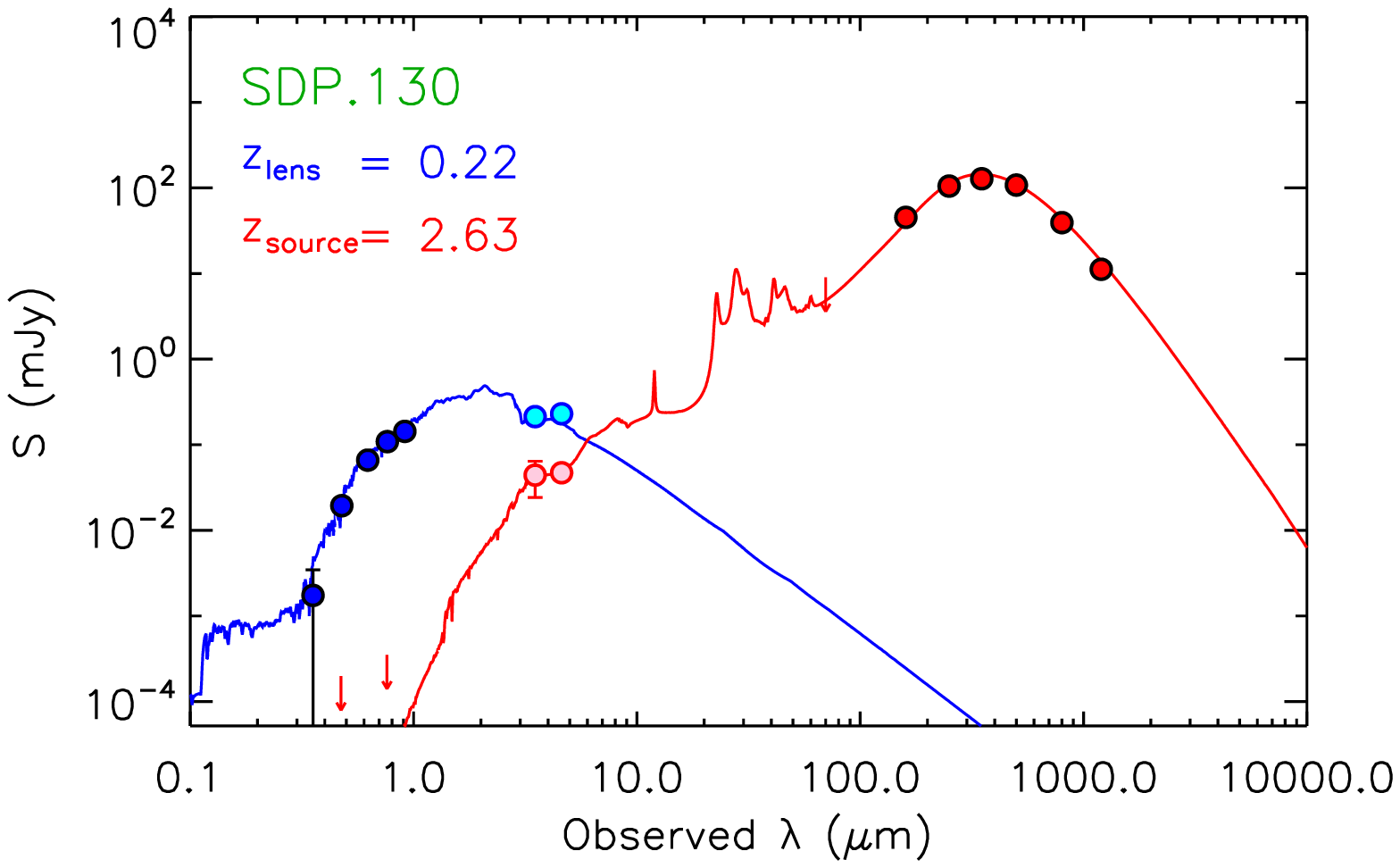}{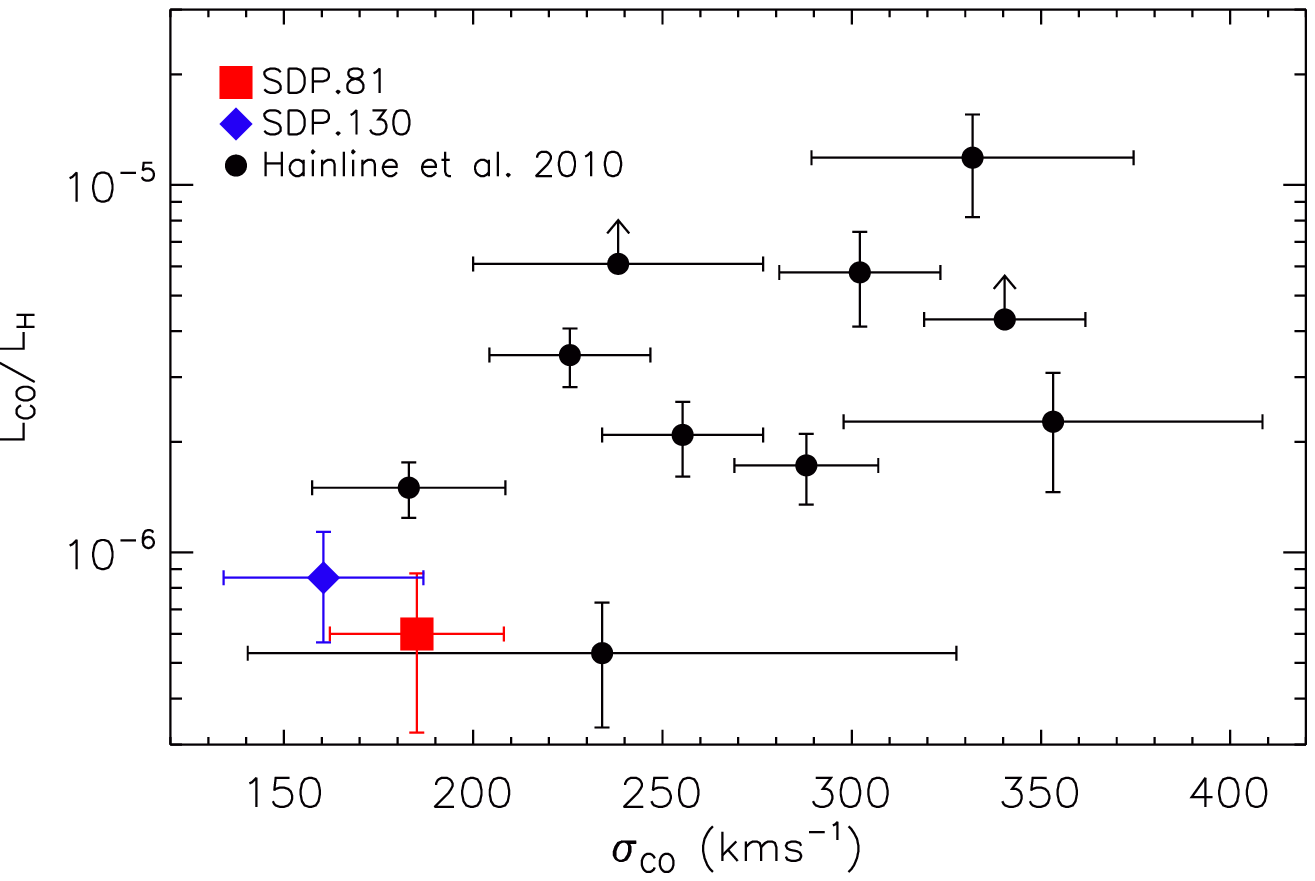}
\end{center}
\vspace{-0.5cm}
\caption{Left: photometry and best-fit SEDs for the foreground elliptical
  (blue) and background SMG (red) for SDP.81 and SDP.130. The
  photometric points and upper limits are taken from Negrello et
  al. (2010), with updated PACS flux densities at 160 $\mu$m and upper
  limits at 70 $\mu$m. Photometry from the best fitting light profiles
  to the IRAC data are added for the foreground lens galaxies
  (turquoise points) and for the background SMG (pink points). The SEDs
  are fitted using the models of da Cunha et al.~(2008). We find that
  high levels of visual extinction of the SMGs are required ($A_{\rm V}
  >$ 4) to be consistent with the optical-to-NIR data. Top right: a plot of (magnification corrected) rest-frame absolute 
$H$-band ($M_{\rm H}$) magnitude against redshift
for SDP.81 and SDP.130, showing the approximate correspondence of $M_{\rm H}$
with stellar mass.  For comparison, we also show 850- and
870-$\mu$m selected galaxies (Hainline et al.~2010; Wardlow et
al.~2010). The two lensed H-ATLAS sources are brighter than $\sim95\%$ of
the comparison sample, and correspondingly, are likely to be amongst
the most massive.  Error bars are omitted from the comparison sample for clarity.
Bottom right: 
ratio of CO and $H$-band luminosity against CO line velocity for the
H-ATLAS lensed galaxies and 850 $\mu$m selected SMGs. $L_{\rm CO}/L_{\rm H}$ is
independent of the lensing model (to the first order; see discussion in the
text) and represents the gas-to-stellar mass ratio. $\sigma_{\rm CO}$
is indicative of dynamical mass, although it also depends on the
inclination angle and size of the CO emitting region, which increases
the observed scatter. SDP.81 and
SDP.130 have small dynamical masses and small gas fractions, relative
to the comparison sample. }
\label{seds}
\end{figure*}

\begin{table}
\scriptsize{
  \caption{\footnotesize Parameters of the Foreground Lens and Background SMG for SDP.81 and SDP.130}
  \label{tab:results}
\begin{center}
 \begin{tabular}{ @{} ccc  @{}}
 \hline
Parameters & SDP.81 & SDP.130\\
 \hline
 \hline
Foreground lens & & \\
\tableline
RA & 09$^{\rm h}03^{\rm m}11.6^{\rm s}$  & 09$^{\rm h}13^{\rm m}05.0^{\rm s}$ \\
Dec & 00$^{\rm d}39^{\rm m}06^{\rm s}$  & $-00^{\rm d}53^{\rm m}43^{\rm s}$\\
Redshift$^{\tablenotemark{a}}$ & 0.299 & 0.220\\
SDSS {\it u}  ($\mu$Jy) & 3.9 $\pm$ 2.0  & 1.7 $\pm$ 1.7\\
SDSS {\it g}   ($\mu$Jy) &  24.9 $\pm$ 1.1  & 19.4 $\pm$ 0.7\\
SDSS {\it r}  ($\mu$Jy) &  115 $\pm$ 2   &  66.1 $\pm$ 1.2\\
SDSS {\it i}  ($\mu$Jy) &  198 $\pm$ 4   & 109 $\pm$ 2\\
SDSS {\it z}  ($\mu$Jy) &  278 $\pm$ 8   & 143 $\pm$ 7\\
UKIDSS {\it Y}  ($\mu$Jy) & 320 $\pm$ 20  & ...\\
UKIDSS {\it J}  ($\mu$Jy) & 370 $\pm$ 20  & ...\\
UKIDSS {\it H}  ($\mu$Jy) & 510 $\pm$ 50  & ...\\
UKIDSS {\it K}  ($\mu$Jy) &  570 $\pm$ 70  & ...\\
{\it Spitzer} 3.6 $\mu$m (mJy) & 0.35 $\pm$ 0.04& 0.213 $\pm$ 0.03 \\
{\it Spitzer} 4.5 $\mu$m (mJy) & 0.22 $\pm$ 0.04 & 0.230 $\pm$ 0.01\\
\tableline
\multicolumn{3}{l}{Background SMG: observed quantities}\\
\tableline
Redshift$^{\tablenotemark{a}}$ & 3.042 & 2.625 \\
Magnification$^{\tablenotemark{a}}$ & 25 $\pm$ 7  & 6 $\pm$ 1 \\
Keck/LRIS {\it g}  ($\mu$Jy) & $<$ 0.13 & $<$ 0.20\\
Keck/LRIS {\it i}   ($\mu$Jy) & $<$ 0.20 & $<$ 0.35\\
UKIDSS {\it Y}   ($\mu$Jy) & $<$ 6.27& ...\\
UKIDSS {\it J}  ($\mu$Jy) &$<$ 9.23 & ...\\
UKIDSS {\it H}   ($\mu$Jy) & $<$ 8.52 & ...\\
UKIDSS {\it K} ($\mu$Jy)  & $<$ 13.5 & ...\\
{\it Spitzer} 3.6 $\mu$m (mJy) &  0.062 $\pm$ 0.04 & 0.044 $\pm$ 0.01\\
{\it Spitzer} 4.5 $\mu$m (mJy) &  0.126 $\pm$ 0.05   & 0.047 $\pm$ 0.01\\
PACS 70 $\mu$m (mJy)$^{\tablenotemark{b}}$   & $<$ 8.0& $<$ 9.0 \\
PACS 100 $\mu$m (mJy) & $<$ 62 & ... \\
PACS 160 $\mu$m (mJy)$^{\tablenotemark{b}}$  & 51 $\pm$ 5 & 45 $\pm$ 8 \\
SPIRE 250 $\mu$m  (mJy) & 130 $\pm$ 20 & 105 $\pm$ 17\\
SPIRE 350 $\mu$m  (mJy) & 180 $\pm$ 30 & 128 $\pm$ 20\\
SPIRE 500 $\mu$m  (mJy) & 170 $\pm$ 30 & 108 $\pm$ 18\\
SMA 880 $\mu$m  (mJy) & 76 $\pm$ 4 & 39 $\pm$ 2\\
IRAM 1200 $\mu$m  (mJy) & 19.6 $\pm$ 0.9 & 11.2 $\pm$ 1.2\\
\tableline
\multicolumn{3}{l}{Background SMG: derived quantities}\\
\tableline
$L_{\rm IR} (10^{12} L_{\odot})^{\tablenotemark{c}}$ &  2.0 $\pm$ 0.6  &  5.6 $\pm$ 1.2 \\
$A_{\rm V}$  &  4.4 $\pm$ 0.6   &  5.0 $\pm$ 0.5 \\
$M_{\rm H}$ & $-25.7\pm0.3$ & $-26.01\pm0.17$ \\
SFR ($M_{\odot}$ yr$^{-1}$)  &  74 $\pm$ 30  &  150 $\pm$ 50\\
$M({\rm H}_2)$ ($10^{10}$ $M_{\odot}$)$^{\tablenotemark{d}}$  &  1.4 & 2.7 \\
$M_{\star}$ ($10^{11}$ $M_{\odot}$)$^{\tablenotemark{e}}$   &  2.5 $\pm$ 1.7 &  4.5 $\pm$ 2.5\\
$M_{\rm dust}$ ($10^{8}$ $M_{\odot}$)$^{\tablenotemark{f}}$   &  3.4 $\pm$ 1.0 & 11 $\pm$ 2 \\
$\mu^{\tablenotemark{g}}$   & 0.05 $\pm$ 0.01 & 0.08 $\pm$ 0.01 \\
 \tableline
\end{tabular}
 \end{center}
\tablecomments{{\scriptsize
The optical and sub-mm flux densities and 3$\sigma$ upper
limits are taken from N10, unless otherwise noted. Derived
 quantities are corrected for magnification
and the errors quoted account for the tabulated uncertainty in the
magnification.
\\ (a) Spectroscopic redshifts (N10). 
\\ (b) From PACS re-imaging data (I. Valtchanov et al.~in preparation 2011; Ibar et
al.~2010). 
\\(c) 2 to 1000 $\mu$m luminosity based on the best-fit SED; 
\\(d) Frayer et al.~2011, uncertainty is at least a factor of $\sim2$. 
\\(e) There is an additional systematic uncertainty up to a factor of
$\sim10$ (see text).
\\(f) Calculated with a dust mass absorption coefficient 
approximated as $\kappa_\lambda \propto \lambda^\beta$ with
$\beta=1.5$ for warm dust and $\beta=2$ for cold dust, and a
 normalization of 0.77 g$^{-1}$ cm$^2$ at 850 $\mu$m. There is 
an additional systematic uncertainty of at least a factor of
$\sim2$.
\\(g) Gas fraction, $\mu = M(H_2)/[M_\star + M(H_2)]$, uncertainty is a
factor of 2--3.}
}
}
\end{table}

\section{Spectral energy distribution of the lensed SMGs}

The GALFIT-integrated magnitudes for each component of the final SDP.81
and SDP.130 models were converted to flux density to extend the
existing multi-waveband photometry (see table \ref{tab:results} and N10, and references therein)
into the near-IR.  We assign 1$\,\sigma$ errors to the photometry obtained
from the final GALFIT model profiles, using the magnitude distributions
for all the GALFIT trials that converged.
PACS re-imaging of the lensed H-ATLAS sources provides new 
photometry at 160$\,\mu$m and upper limits at
70$\,\mu$m (I. Valtchanov et al.~in preparation; Ibar et al.~2010). For the goal of
deriving physical properties, the IRAC photometry adds particularly
important constraints to the SMG SEDs, which previously consisted of
just upper limits at wavelengths below 250$\,\mu$m for SDP.130 and
160$\,\mu$m for SDP.81.

The SEDs of the SMGs are fitted using the models of da Cunha et
al.~(2008), calibrated to reproduce the ultraviolet-to-infrared SEDs of
local, purely star-forming Ultra Luminous Infrared Galaxies (ULIRGs;
10$^{12} \leq$ $L_{\rm IR}/L_{\odot} < 10^{13}$; da Cunha et al.~2010).
The SED models assume a Chabrier (2003) initial mass function (IMF) that is cutoff below 0.1
and above 100 $M_{\odot}$; using a Salpeter IMF instead gives stellar
masses that are a factor of $\sim1.5$ larger. We find that a
significant attenuation by dust ($A_{\rm V} \sim $4-5) is required to
be consistent with the IRAC photometry and optical/near-IR upper limits
(Figure~\ref{seds}), which is consistent with other ULIRGs and SMGs (e.g.
Geach et al.~2007; Hainline et al.~2010; Micha{\l}owski et al.~2010;
Wardlow et al.~2010). 

Using the Chabrier (2003) IMF, with parameters derived from the SED
fits, we find that SDP.81 and SDP.130 have stellar masses ($M_\star$)
of $(2.5\pm1.7)\times10^{11}$~$M_{\odot}$ and
$(4.5\pm2.5)\times10^{11}$~$M_{\odot}$, respectively (see table \ref{tab:results}). However, we note
that there is an additional systematic error of up to a factor of 10,
due to uncertainty in the appropriate mass-to-light ratio (see Wardlow
et al. 2010 for a discussion) and magnification factors, as well as a lack of
observations at optical/near-IR wavelengths.  {\it J} and {\it K}s photometric data
on these galaxies with VLT/HAWK-I observations (A. Verma et al. in preparation 2011)
could potentially improve the estimates of extinction and stellar mass.

Rest-frame absolute {\it H}-band magnitudes ($M_{\rm H}$) provide a guide to
galaxy stellar masses that is not complicated by the details of the
assumed star formation history and is straightforward to compare to
other similarly selected samples.  In the top panel of Figure~\ref{seds} we plot $M_{\rm H}$
against redshift for SDP.81 and SDP.130 compared to 850 and 870 $\mu$m
selected SMGs (Hainline et al.~2010; Wardlow et al.~2010). The H-ATLAS
lensed galaxies are brighter than the majority of the $850$--$870$ $\mu$m
selected SMGs, thus if there is no active galactic nucleus contribution to their $H$-band
luminosities they are likely to be amongst the most massive.

We next consider the CO(1$-$0)-to-$H$-band luminosity ratio ($L_{\rm CO}/L_{\rm H}$), a
representation of the gas-to-stellar mass fraction that is mostly
independent of the lensing model. However, the IRAC data
trace stellar emission, which may originate from a spatially different
region of the SMG to the far-infrared emission. As such there may
be variation in the lensing amplifications across the source plane,
potentially leading to a small effect on $L_{\rm CO}/L_{\rm H}$
 The bottom panel of Figure~\ref{seds}
shows $L_{\rm CO}/L_{\rm H}$ against the CO linewidth ($\sigma_{CO}$) for the two
H-ATLAS lensed galaxies, compared to 850-$\mu$m selected SMGs (Hainline
et al.~2010). $\sigma_{\rm CO}$ is independent of $L_{\rm CO}/L_{\rm H}$ and is
indicative of the dynamical mass of the system, although some scatter
is introduced due to the dependence on the inclination angle and the
size of the CO emitting region.  Published CO luminosities
of the 850 $\mu$m selected sample (Greve et al.~2005, Coppin
et al.~2008; Frayer et al.~2008, Tacconi et al.~2008; Bothwell et
al.~2010) are converted to the equivalent CO(1$-$0) values using
$L_{CO(3-2)}/L_{CO(1-0)}=0.6$ and $L_{{\rm CO}(4-3)}/L_{{\rm CO}(1-0)}=0.6$ (Ivison
et al.~2010). Galaxies with poorly defined $\sigma_{{\rm CO}}$ are excluded.
The H-ATLAS lensed galaxies have low $L_{{\rm CO}}/L_H$
(representative of gas fraction) and low $\sigma_{{\rm CO}}$ (representative
of dynamical mass) compared to 850-$\mu$m selected galaxies
(Figure~\ref{seds}). There may also be a weak inverse trend between
$L_{{\rm CO}}/L_{\rm H}$ and $\sigma_{{\rm CO}}$ for all the SMGs, although a larger
sample is required for confirmation. 

\vspace*{0.4cm} 
{\it Summary}: We have studied the background SMGs of two sub-mm bright
gravitational lenses, identified in the H-ATLAS SDP data. The intrinsic
sub-mm flux densities of these SMG are below the {\it Herschel}
confusion noise and, therefore, undetectable without the fortuitous
lensing by foreground ellipticals.  The full 550 deg$^2$ survey area
of H-ATLAS will recover a sample of $>200$ lensed SMGs, and {\it
  Spitzer} follow-up observations will enable us to study the physical
properties of SMGs over a wide range of redshift and far-IR
luminosities. Studies such as this are necessary to further understand
the nature of sources that contribute to the bulk of the cosmic
far-infrared background.

\vspace{-0.5cm}
\acknowledgments

{\it Herschel}-ATLAS is a project with {\it Herschel}, which is an ESA
space observatory with science instruments provided by European-led
Principal Investigator consortia and with important participation from
NASA. The H-ATLAS Web site is http://www.h-atlas.org/.  We thank the
Science and Technology Facilities Council, grant D/002400/1 and
studentship SF/F005288/1.  This work is based in part on observations
made with the {\it Spitzer} Space Telescope, which is operated by the
Jet Propulsion Laboratory, California Institute of Technology under a
contract with NASA.  Support for this work was provided by NASA through
an award issued by JPL/Caltech.  US participants in H-ATLAS also
acknowledge support from NASA {\it Herschel} Science Center through a
contract from JPL/Caltech.  IA and DHH are partially funded by CONACyT
grants 50786 and 60878.


\begin{thebibliography}{}

\bibitem{Blain} Blain, A. W. 1996, \mnras, 283, 1340

\bibitem{Bothwell} Bothwell, M. S., et al. 2010,
  \mnras, 405, 219

\bibitem{Chabrier} Chabrier, G. 2003, \pasp, 115, 763

\bibitem{Coppin} Coppin, K., et al., 2008, \mnras, 389, 45

\bibitem{da Cunha1} da Cunha, E., Charlot, S., \& Elbaz, D. 2008, \mnras, 388, 1595

\bibitem{da Cunha} da Cunha, E., Charmandaris, V., D\'{i}az-Santos, T., Armus, L., Marshall, J. A., \& Elbaz, D. 2010, \aap, 523, A78

\bibitem{de Vaucouleurs} de Vaucouleurs, G. 1948, Ann. Astrophys., 11, 247

\bibitem{Eales} Eales, S., et al. 2010, \pasp, 122, 499 

\bibitem{Fazio} Fazio, G. G., et al. 2004, \apjs, 154, 10 

\bibitem{Frayer1} Frayer, D. T., et al. 2008, \apj, 680, L21

\bibitem{Frayer2} Frayer, D., et al. 2011, \apjl, 726, L22

\bibitem{Geach} Geach, J. E., Smail, I., Chapman, S. C., Alexander, D. M., Blain, A. W., Stott, J. P., \& Ivison, .R. J. 2007, \apj, 655, L9

\bibitem{Greve} Greve, T. R., et al. 2005, \mnras, 259, 1165

\bibitem{Hainline} Hainline, L. J., Blain, A. W., Smail, I., Alexander, D. M., Armus, L., Chapman, S. C., \& Ivison, R. J. 2010, \mnras, submitted (arXiv:1006.0238)

\bibitem{Ibar} Ibar, E., et al. 2010, \mnras, 409, 38

\bibitem{Ivison} Ivison, R. J., Papadopoulos, P. P., Smail I., Greve, T. R., Thomson, A. P. , Xilouris, E. M., \& Chapman, S. C. 2010, \mnras, in press (arXiv:1009.0749)

\bibitem{Lupu} Lupu, R., et al. 2010, \apj, submitted (arXiv:1009.5983)

\bibitem{Makovoz} Makovoz, D., $\&$ Marleau, F. R. 2005, \pasp, 117, 1113

\bibitem{McCarthy} McCarthy, J. K., et al. 1998,  Proc. SPIE, 3355, 81

\bibitem{Michalowski} Micha{\l}owski, M. J., Hjorth, J., \& Watson, D. 2010, A\&A, 514, A67

\bibitem{Negrello} Negrello, M., Perrotta, F., Gonz\'{a}lez-Nuevo, J., Silva, L., de Zotti, G, Granato, G. L., Baccigalupi, C. \& Danese, L. 2007, \mnras, 377, 1557

\bibitem{Negrello2} Negrello, M., et al. 2010, Science, 330, 800

\bibitem{oke95} Oke, J. B., et al., 1995, \pasp, 107, 375

\bibitem{Pascale} Pascale, E., et al. 2010, \mnras, submitted (arXiv:1010.5782)

\bibitem{Peng} Peng, C. Y., Ho, L. C., Impey, C. D., $\&$ Rix, H.-W. 2002, \aj, 124, 266

\bibitem{Perrotta} Perrotta, F., Baccigalupi, C., Bartelmann, M., De Zotti, G. $\&$ Granato, G.~L.  2002, \mnras, 329, 445

\bibitem{Pilbratt} Pilbratt, G. L., et al. 2010, A\&A, 518, L1

\bibitem{Rigby} Rigby, E. E., et al. 2010, \mnras, submitted (arXiv:1010.5787)

\bibitem{Tacconi} Tacconi, L. J., et al. 2008, \apj,
  680, 246

\bibitem{Treu} Treu, T. 2010, \araa, 48, 87

\bibitem{Wardlow} Wardlow, J. L., et al. 2010, \mnras,
  submitted, (arXiv:1006.2137)

\end{thebibliography}
\end{document}